\documentclass[aps,prl,twocolumn,groupedaddress]{revtex4}

\usepackage{graphicx}
\usepackage{epstopdf}
\usepackage{natbib}
\usepackage{hyperref}

\begin{document}


\title{Reversing the temporal envelope of a heralded single photon using a cavity}

\author{Bharath Srivathsan}
\altaffiliation{Center for Quantum Technologies, 3 Science Drive 2, Singapore  117543}
\author{Gurpreet Kaur Gulati}
\altaffiliation{Center for Quantum Technologies, 3 Science Drive 2, Singapore  117543}
\author{Alessandro Cer\`{e}}
\altaffiliation{Center for Quantum Technologies, 3 Science Drive 2, Singapore  117543}
\author{Brenda Chng}
\altaffiliation{Center for Quantum Technologies, 3 Science Drive 2, Singapore  117543}
\author{Christian Kurtsiefer}
\altaffiliation{Center for Quantum Technologies and Department of Physics,
  National University of Singapore, 3 Science Drive 2, Singapore  117543}
\email[]{christian.kurtsiefer@gmail.com}
\date{\today}
\begin{abstract}
We demonstrate a way to prepare single photons with a temporal envelope that resembles the time reversal of photons from the spontaneous decay process. We use the photon pairs generated from a time-ordered atomic cascade decay:
the detection of the first photon of the cascade is used as a herald
for the ground-state transition resonant second photon.
We show how the interaction of the heralding photon with an asymmetric Fabry-Perot cavity
reverses the temporal shape of its twin photon from a decaying to a rising exponential envelope.
This single photon is expected to be ideal for interacting with two level systems.
\end{abstract}


\maketitle
Absorption of a single photon by a single atom or an ensemble of atoms is an interesting problem from
a fundamental point of view, and is also essential for many quantum information protocols~\cite{cirac:1997,kimble:2008,DLCZ:2001}.
One of the requirements for an efficient absorption is that the temporal shape of the incident photon 
is the time reversal of the photon from the spontaneous decay process~\cite{Stobinska:2009,Yimin:2011}.
Temporally shaped light pulses have been utilized in many recent experiments to achieve efficient interactions between light and matter~\cite{Ritter:2012,Lvovsky:2009}.
In particular, the advantage of using a rising exponential shaped single photon for absorption in an atomic ensemble was demonstrated in~\cite{Zhang:2012}, and shaped multiphoton pulses for exciting a single atom was demonstrated in~\cite{Syed:2013}.

Efficient preparation of single photons with
narrow bandwidth and
a rising exponential envelope is not trivial.
One solution is the direct modulation of a heralded photon generated by an atomic medium~\cite{Kolchin:2008}.
This technique results in unavoidable losses due to filtering.
We have previously demonstrated a scheme to generate single photons with a rising exponential shape 
by heralding on photon pairs produced by cascade decay~\cite{Gulati:2014}
without filtering.
The drawback of this scheme is that the photon with the rising exponential envelope is
not resonant with an atomic ground state transition.

In this letter, we combine the asymmetric cavity design used by Bader et al.~\cite{Bader:2013}
with the well known temporal correlation properties of photon pairs~\cite{Franson:1992cl}
to invert the temporal envelope of the generated photon pairs:
with the proper heralding sequence we obtain a rising exponential single photon resonant with 
a ground state transition of $^{87}$Rb.

The photons emerging from an atomic cascade decay have a well defined time
order. The first photon of the cascade (signal) is generated before the photon
resonant with the ground state (idler). 
The resulting state can be described by a two photon wave function \cite{Jen:2012} of the form
\begin{equation}
  \label{eq:wavepacket}
  \psi(t_{s},t_{i})=A\,e^{-\,(t_{i}-t_{s})/2\,\tau}\,\Theta(t_{i}-t_{s})\,,
\end{equation}
where $t_{s}, t_{i}$ are the detection times of the signal and idler
photons, and $\Theta$ is the Heaviside step function. In this notation, the
probability of observing a pair is proportional to $\left|\psi(t_{s},t_{i})\right|^2$. 
The exponential envelope and the decay time $\tau$ is
a consequence of the atomic evolution of the cascade decay.
If the detection of a signal photon is used as herald, the idler mode has a
single photon state with a exponentially decaying temporal envelope starting
at $t_i=t_s$.
Similarly, if the detection of an idler photon acts as a herald,
the signal photon has an exponentially rising temporal envelope.

An asymmetric cavity with the appropriate parameters transforms a light field with an exponentially rising envelope into one
with a decaying envelope~\cite{Bader:2013}.
The main point of this work is that
coupling the signal mode into a properly tuned asymmetric cavity 
before heralding results in a ``time reversed'' envelope for the idler photon.

\begin{figure}
  \begin{center}
    \includegraphics[width=\columnwidth]{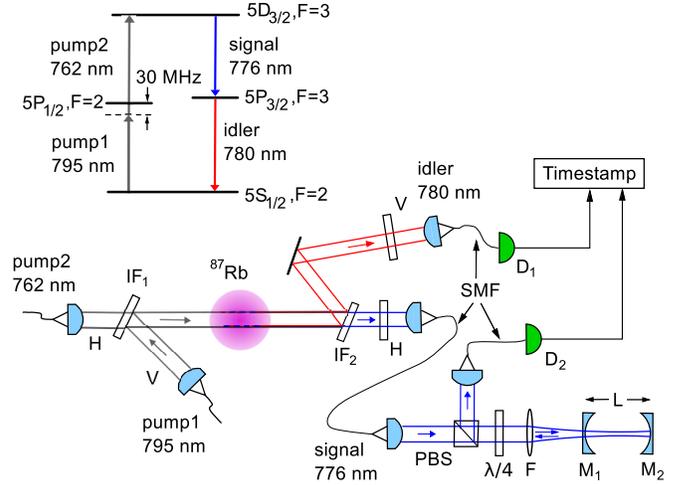}
   \caption{\label{fig:schematic}Schematic of the four-wave mixing experiment in
    collinear geometry. IF$_1$, IF$_2$: interference filters, used to combine pump beams
    and to separate the photons pairs. SMF: Single mode optical fibers. M$_1$,
    M$_2$: cavity mirrors.
    The incoming and outgoing mode
    of the cavity are separated by a polarising beam splitter (PBS) and a quarter wave plate
    ($\lambda/4$). 
    \mbox{D$_1$, D$_2$}: silicon avalanche photodiodes (APD).
    The inset shows the cascade level scheme for generation of
    photon pairs in $^{87}$Rb.}
  \end{center}
\end{figure}

The asymmetric cavity is formed by a partially
reflective mirror M$_1$, and a highly reflective mirror M$_2$, see Fig.~\ref{fig:schematic}.
The effect of the cavity on the signal mode is described as a frequency-dependent phase
factor~\cite{hodgson:1997},
\begin{equation}\label{eq:transformation}
	C(\delta')\,=\,
  \frac{\sqrt{R_1}\,-\,
  \,\sqrt{R_2}\,e^{i\,\delta'/\Delta\nu_f}}{1 - 
  \,\sqrt{R_1\,R_2}\,e^{i\,\delta'/\Delta\nu_f}}\,,
\end{equation}
where $R_{1,2}$ are the reflectivities of M$_1$ and M$_2$, $\Delta\nu_f$ is the free
spectral range of the cavity, and $\delta'$ the detuning from the cavity resonance.
For $R_2=1$, the transformation of the incoming mode is lossless, i.e.,
$|C(\delta')|=1$.

The cavity transforms the two-photon wavefunction in Eq.~(\ref{eq:wavepacket})
into the two-photon wavefunction $\tilde{\psi}(t_{s}, t_{i})$:
\begin{equation}\label{eq:tr2}
  \tilde{\psi}(t_{s},t_{i})=\mathcal{F}_s^{-1}\left[
     C(\omega_s-\omega_s^0-\delta) \cdot 
     \mathcal{F}_s\left[\psi(t_s, t_i)\right]
   \right]\,,
\end{equation}
where $\mathcal{F}_s$ denotes a Fourier transform from $t_s$ to  $\omega_s$,
and $\delta$ is the detuning of the cavity resonance from the signal
photon center frequency $\omega^0_s/2\pi$.

If the ring-down time of the cavity matches the coherence time $\tau$ of the
photon pair in Eq.~(\ref{eq:wavepacket}),
the resulting wavefunction is:
\begin{eqnarray}\label{eq:wavepacket1}\nonumber
	\tilde{\psi}(t_{s},t_{i}) =
		\frac{A}{\sqrt{1+4\,\delta^2\tau^2}}\,
		[&2\,\delta\;\tau\,e^{-(t_i-t_{s})/2\,\tau}\,\Theta(t_{i}-t_{s})\,\\
		&+ e^{(t_i-t_{s})/2\,\tau}\,\Theta(-t_{i}+t_{s})]
\end{eqnarray}
with an exponentially rising and an exponentially
decaying component. Their
relative
weight can be controlled by the detuning
$\delta$, and for $\delta\,=\,0$, a time-reversed version of
Eq.~(\ref{eq:wavepacket}) is obtained:  
\begin{equation}
\label{eq:wavepacket2}
\tilde{\psi}(t_{s},t_{i})\,=\,A\,e^{(t_{i}-t_{s})/2\,\tau}\,\Theta(-t_{i}+t_{s}).
\end{equation}

Heralding on the detection of 
the modified signal photon results in an 
idler photon state with a rising exponential
envelope, ending at $t_i=t_{s}$.
The cavity thus effects a reversal of the
temporal envelope of the heralded idler photons from an exponential decay to a
rise. 

To experimentally investigate this method, we used the setup shown in
Fig.~\ref{fig:schematic}. We generate time-ordered photon pairs by
four-wave mixing in a cold ensemble of $^{87}$Rb atoms in a cascade level
scheme. Pump beams at 795\,nm and 762\,nm excite atoms from the $5S_{1/2},F=2$
ground level to the $5D_{3/2},F=3$ level via a two-photon transition. The 776\,nm
(signal) and 780\,nm (idler) photon pairs emerge from a cascade decay back to
the ground level and are coupled to single mode fibers.
All four modes are collinear and propagate in the same direction.
The coherence time $\tau$ of the 
photon pairs is determined by a time-resolved coincidence measurement between
the  detection of signal and idler photons to be 5.9\,ns. Details about the
photon pair source can be found in~\cite{Srivathsan:2013}
and~\cite{Gulati:2014}.

One of the modes of the photon pairs (signal in Fig.~\ref{fig:schematic}) is
coupled to the fundamental transverse mode of an asymmetric cavity, formed by
mirrors M$_1$, M$_2$ with  radii of curvature of 100\,mm and 200\,mm,
respectively. We characterize the cavity using a frequency stabilized laser of 
wavelength 776\,nm. The reflectivity of mirror M$_1$ is determined by 
direct measurement with a PIN photodiode to be $R_1=0.9410\pm0.0008$. 
Transmission through the mirror M$_2$ and absorption by the mirrors 
leads to losses in the cavity. The loss per round trip is 
determined from the transmission through and the reflection from the cavity 
and is included in the reflectivity of M$_2$, $R_2=0.998\pm0.001$.
The mirrors are separated by $5.5$\,cm corresponding to a 
free spectral range $\Delta\nu_f=2.7$\,GHz. Therefore, an incident photon of Fourier 
bandwidth $1/(2\pi\tau)=27$\,MHz interacts effectively with only one
longitudinal mode of the cavity, ensuring that Eq.~(\ref{eq:transformation}) is an
adequate model. The light reflected off the cavity is separated from
the incident mode by using a polarising beam splitter (PBS) and a quarter waveplate
($\lambda/4$).

\begin{figure}
  \begin{center}
    \includegraphics[width=\columnwidth]{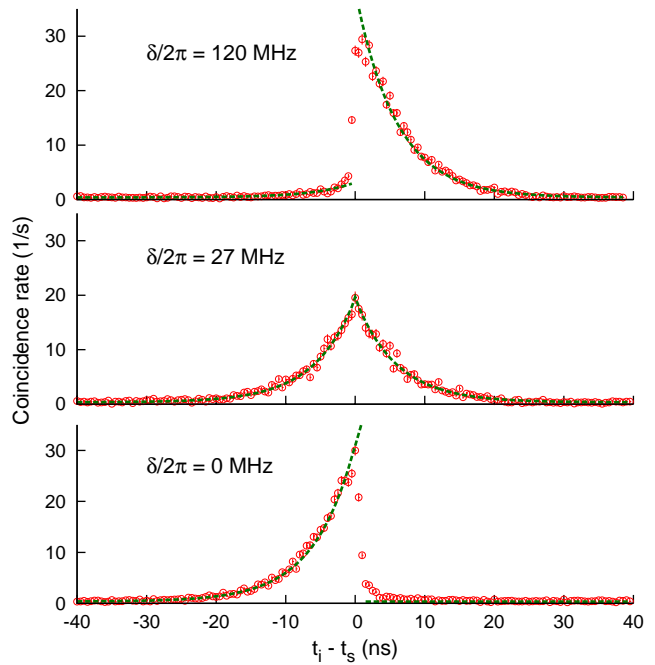}
    \caption{\label{fig:plot1}
      Transformation of the temporal envelope of the heralded idler photon
      from exponential decay to rise when the cavity is in the signal mode.
      The y-axis shows the coincidence rate $G_{si}^{(2)}$ between the detectors D$_1$ and D$_2$ as a function of the detection time difference.
      The dashed lines represent $\left|\tilde{\psi}(t_{s},t_{i})\right|^2$,
      obtained from the model described by Eq.~(\ref{eq:wavepacket1}) for the
      indicated cavity detunings $\delta$,
      with amplitude $A$ as the only free parameter used to fit the experimental points.}
  \end{center}
\end{figure}

We infer the temporal shape of the heralded photons
from the time distribution of the coincidence rate $G_{si}^{(2)}$ between the APDs D$_1$ and D$_2$ (time resolution $<1$\,ns).
In Fig.~\ref{fig:plot1}
we show
$G_{si}^{(2)}$ for three different cavity-photon detunings.
When the cavity resonance is tuned
far away from from the signal photon frequency $\omega_s^0$,
in this case about $\delta/2\pi=120$\,MHz,
the temporal envelope remains nearly unchanged
from the exponential decay obtained without cavity.
At $\delta/2\pi=27$\,MHz, 
the time distribution becomes a symmetric exponential, 
and on resonance, $\delta/2\pi=0$,
we obtain a rising exponential shape.
For the three detunings 
the measurement agrees with the shape expected from Eq.~(\ref{eq:wavepacket1}):
the exponential time constants remain unchanged
and the new temporal shapes are determined by the phase shift across the cavity resonance via Eq.~(\ref{eq:transformation}).

From the time distribution of the coincidence counts
it is evident that
the situation is
symmetrical to what we presented in~\cite{Gulati:2014}:
by heralding on the signal photon
we now obtain an idler photon with a rising exponential temporal envelope.
This result, though predicted by the theory, is particularly exciting:
the idler photon is resonant with a ground state transition
and the obtained temporal envelope is 
similar to
the time reversal of the one obtained by spontaneous decay.
The only deviation from the predicted shape occurs for a short
time interval after the detection of the herald ($t_i-t_s > 0$).
We attribute this deviation to an imperfect matching between the signal and cavity spatial modes.
\begin{figure}
  \begin{center}
    \includegraphics[width=\columnwidth]{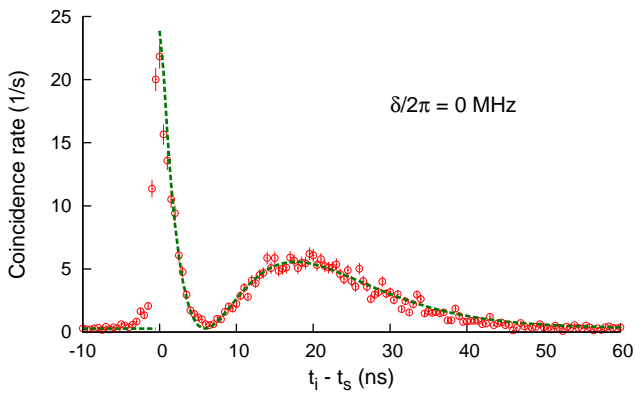}
    \caption{\label{fig:plot2} 
    Temporal envelope of the idler photon when the cavity is aligned and tuned to resonance with the idler mode.
    The dashed line represents
    $\left|\tilde{\psi}(t_{s},t_{i})\right|^2$
    obtained from Eq.~(\ref{eq:tr2}) by swapping $i$ and $s$.
    Also in this case, the amplitude is the only free parameters used in the fit.}
  \end{center}
\end{figure}

To confirm the predictive power of our model,
we repeated the same experiment swapping the roles of the signal and idler modes.
This corresponds to swapping the subscripts $s$ and $i$
in Eqs.~(\ref{eq:tr2}) and~(\ref{eq:wavepacket1}).
Figure~\ref{fig:plot2} shows the 
time resolved coincidence rate $G_{si}^{(2)}$ between the signal 
and modified idler photons
with the cavity tuned to resonance with the idler central frequency.
In this case the cavity transforms the exponentially rising temporal envelope
into a more complex shape.
Our model describes accurately this complex shape, as can be seen from the dashed line in Fig.~\ref{fig:plot2}.

Using the same setup,
we can infer the population of the cavity mode as a function of time, 
and observe its dependence on the envelope of the incident photon. 
\begin{figure}
\begin{center}
  \includegraphics[width=\columnwidth]{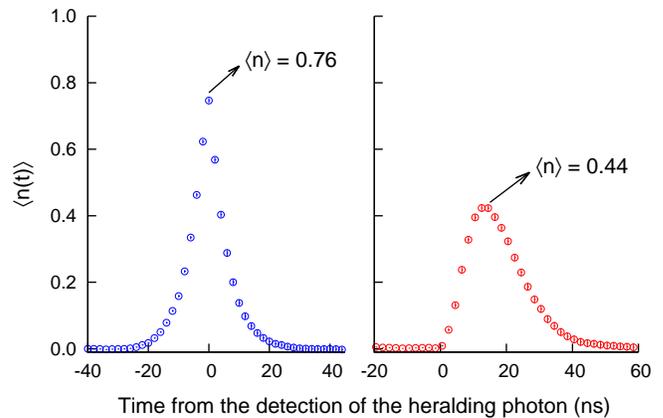}
  \caption{\label{fig:excitation} Mean photon number in the cavity
  estimated using Eq.~(\ref{eq:n_mean}).
  On the left, the detection of an idler photon is used as herald and the cavity is in the signal mode.
  In this case we observe the interaction of an exponentially rising waveform with the cavity.
  On the right, the roles of signal and idler are swapped and the cavity interacts with an exponentially decaying incident photon.
   }
\end{center}
\end{figure}
We estimate the mean photon number $\langle n(t) \rangle$ in the cavity
using an algorithm similar to the one presented in~\cite{Bader:2013}.
We compare the time distributions of coincidence rates
$G_{fr}^{(2)}$ and $G_{or}^{(2)}$
when the cavity is tuned far-off resonance ($\delta/2\pi=200$\,MHz) and on resonance ($\delta/2\pi=0$\,MHz) 
with the incident photons, normalized against the total far-off resonance
coincidences. 
\begin{equation}\label{eq:n_mean}
  \langle n(t)\rangle\,=\,
  \frac{
  e^{-\eta t \Delta\nu_f}
  \int\limits_{-\infty}^{t} \left[G_{fr}^{(2)}(t') -  G_{or}^{(2)}(t') \right] e^{\eta t' \Delta\nu_f}\mathrm{d}t'
  }
  {
  \int \limits_{-\infty}^{\infty} G_{fr}^{(2)}(t') \mathrm{d}t'
  }.
\end{equation}
We estimated $\eta$, that includes the cavity losses per round trip and
transmission through M$_2$, to be $0.002\pm0.001$.

When the cavity is exposed to the idler mode,
a  heralded single photon with decaying exponential envelope 
interacts with the cavity:
the mean photon number in the cavity reaches a maximum of $0.44\pm0.01$.
On the other hand, when the cavity is aligned in the signal mode we have a
heralded single photon with an increasing exponential envelope interacting with
the cavity;
in this case, $\langle n(t)\rangle$ reaches a maximum of $0.76\pm0.01$.
As expected, the photon with the rising exponential waveform interacts more efficiently with the cavity.
Following the analogy in~\cite{Heugel:2010}, we expect this result to be extended to the probability of absorption of a single photon by a single atom. 
In the case of interaction with a single atom, 
it will be necessary to also match the bandwidth of the transition.
We have already demonstrated how it is possible to control the bandwidth of the photon generated by the cascade process by adjusting the optical density of the atomic medium~\cite{Srivathsan:2013}.

In summary, we have demonstrated a method to transform a heralded single
photon with a decaying exponential temporal envelope to a rising exponential
envelope using a cavity. Using this method, we obtain single photons that
resembles the time-reversed versions of photons from  spontaneous decay
process resonant to the D2 line of $^{87}$Rb atoms. Single photon states of
this envelope and bandwidth would be useful for transferring information from
photons back into atoms, completing the toolbox necessary to interconnect
atoms in a complex quantum information processing scenario.
  
We acknowledge the support of this work by the National Research Foundation \&
Ministry of Education in Singapore.


\end{document}